\input{aipcheck}

\documentclass[
    ,final            
  ]
  {aipproc}

\usepackage{AASstyles}
\layoutstyle{8x11single}

\begin{document}

\title{Spectral evolution and the onset of the X-ray GRB afterglow}


\author{P.A. Evans}{
  address={X-ray and Observational Astronomy Group, Dept. of Physics \&\
Astronomy, University of Leicester, Leicester, UK, LE1 7RH}
}

\author{J.P. Osborne}{}
\author{R. Willingale}{}
\author{P.T. O'Brien}{}

\begin{abstract}
Based on light curves from the Swift Burst Analyser, we investigate 
whether a `dip' feature commonly seen in the early-time hardness ratios
of Swift-XRT data could arise from the juxtaposition of the decaying
prompt emission and rising afterglow. We are able to model the dip as
such a feature, assuming the afterglow rises as predicted by
\citet{Sari1999}. Using this model we measure the initial bulk
Lorentz factor of the fireball. For a sample of 23 GRBs we find a median
value of $\Gamma_0$ = 225, assuming a constant-density circumburst
medium; or  $\Gamma_0$ = 93 if we assume a wind-like medium.
\end{abstract}

\maketitle


\section{Introduction} The early Swift-XRT observations of GRBs have
long been seen to exhibit spectral evolution (\cite{Starling2005};
\cite{Butler2007}). A feature seen in a number of GRBs is a `dip' in the
hardness ratio (i.e. a softening followed by a hardening; see Fig.~1a)
at the end of the steep decay phase. The Swift Burst Analyser light
curves (\cite{BurstAn}), which are in units of the intrinsic (i.e.\
unabsorbed) flux and account for spectral evolution, suggest that
this may correspond to the `turn-on' on the X-ray afterglow (Fig.~1b).

The rise of the afterglow was considered by \cite{Sari1999}, who
predicted that it rise as t$^2$ until some time $t_p$ at which point it
peaks and begins to decay. This time $t_p$ corresponds to the
deceleration radius: the radius at which the rest mass energy of the
material swept up by the fireball equals the  energy of the fireball.
From this one can determine the bulk Lorentz factor at the deceleration
radius of the fireball (e.g.\ \cite{Sari1999}; \cite{Molinari2007}).

\begin{equation}
\frac{\Gamma_0}{2} = \left(\frac{3E_{\rm iso} \left[1+z\right]^3} {32\pi n
m_p c^5\eta t^3_p}\right)^{\frac{1}{8}}
\end{equation}

Where $E_{\rm iso}$ is the isotropic equivalent energy output of the
GRB, $n$ is the number density of the circumburst medium, and $\eta$ is
the radiative efficiency of the fireball.

This assumes that the density is constant, i.e. the GRB is in an
ISM-like environment. If we instead assume that the environment is
wind-like (i.e.\ $\rho\propto r^{-2}$) we find:

\begin{equation}
\frac{\Gamma_0}{2} = \left(\frac{E_{\rm iso} \left[1+z\right]} {8\pi A
m_p c^3\eta t_p}\right)^{\frac{1}{4}}
\end{equation}

where $A$ is the density normalisation.  Following Molinari et
al.\ we assume $\eta=0.2$, $n=1$ cm$^{-3}$ and $A=3\times10^{35}$
cm$^{-1}$; noting that dependence on these unknowns is weak.

In this paper we will therefore consider whether the X-ray hardness ratio dip
can be used to constrain the rise of the X-ray afterglow and hence the
initial bulk Lorentz factor, given the Sari \&\ Piran model. It should
be noted that the presence of a plateau phase in X-ray data is often
interpreted as a sign of continued energy injection into the fireball in
some form or other (e.g. \cite{Zhang2006}). It is likely that this
will modify the fireball dynamics somewhat, however since the rate of
energy injection is much lower than the rate of energy output in the
prompt GRB, we believe we can safely disregard this effect.

\begin{figure}
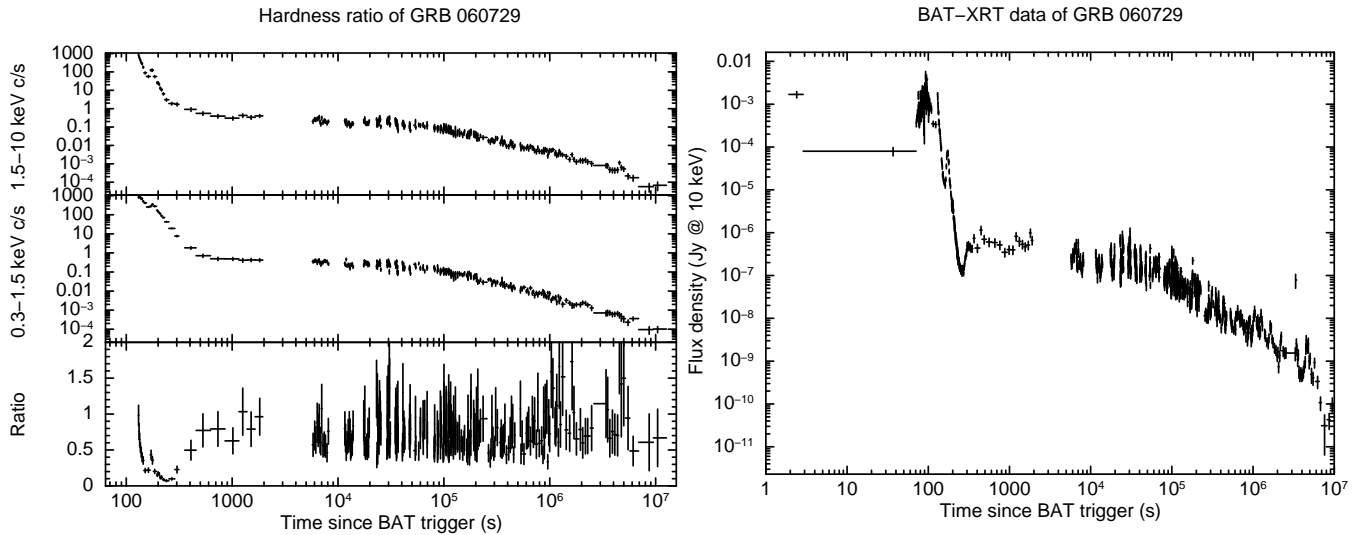

\hbox{
  \includegraphics[height=.4\textheight,angle=-90]{GRB060729_hr.ps}
  \includegraphics[height=.4\textheight,angle=-90]{GRB060729_bann.ps}
}
\caption{The (1.5--10)/(0.3--1.5) keV hardness ratio of GRB 060729 (left
panel), showing the `dip' feature around 100--200 s after the trigger.
The reconstructed 10 keV flux density light curve from the Burst
Analyser light curve (right panel) -- which accounts for this spectral
evolution -- shows a feature at this same time, suggestive of the
overlap of the decaying prompt emission and rising afterglow.}
\end{figure}

\section{Model fitting} 

To investigate whether the hardness ratio dip can indeed be caused by
the rise of the afterglow, we fit the hard (1.5--10 keV) and soft
(0.3--10 keV) XRT band data from the XRT light curve repository
(\cite{LightCurves}, \cite{Catalogue}) simultaneously. We fit two
components: the first models the decaying prompt emission which we
parameterise with the pulse model described by Willingale et al.\
(2005); the hardness ratio of this component decreases (i.e. softens) as
a power-law. The second component model is the afterglow, which we treat
as spectrally invariant. This component is zero until some time $t_a$ at
which the afterglow `starts', thereafter it rises as $t^2$ until $t_p$
whereafter it breaks to a generic $t^{-\alpha}$ decay. 

We do not fit the entire light curve, since we are only interested in
the transition from the prompt to afterglow emission. If the afterglow
light curve shows evidence for a further break after that referred to
above, the data after this break are excluded from the fit. Similarly
any breaks or flares which occur before the prompt emission follows a
single power-law are excluded; if the burst shows none of these features
the entire dataset is fitted. Example fits are shown in Fig.~2.
\begin{figure}
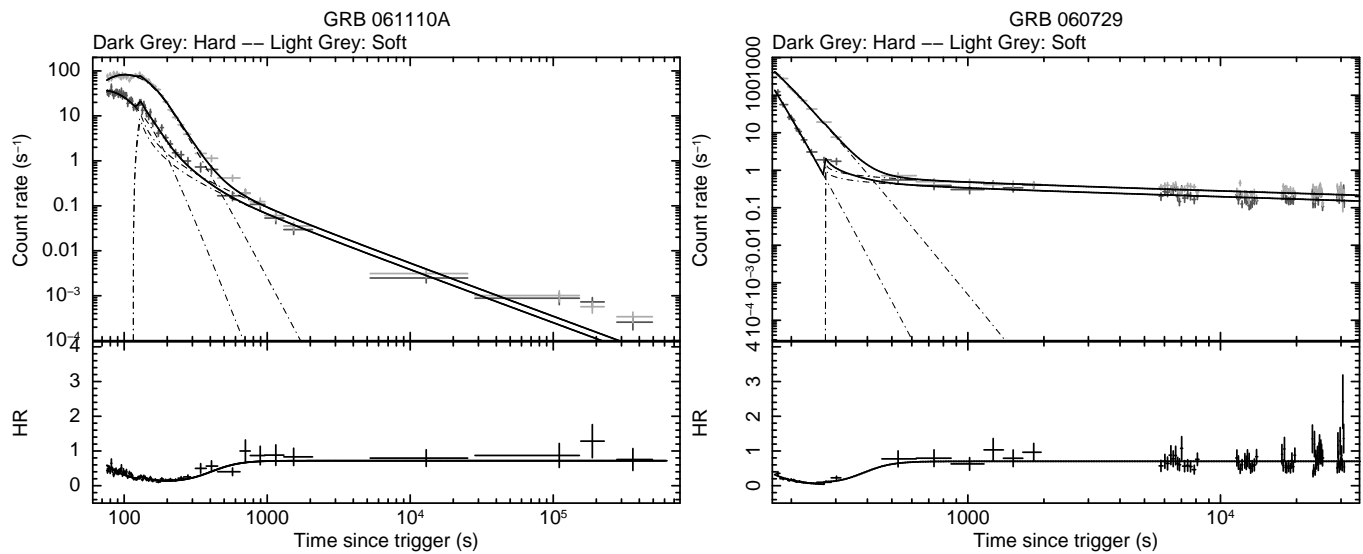

\hbox{
  \includegraphics[height=.4\textheight,angle=-90]{conf_fit_061110A.eps}
  \includegraphics[height=.4\textheight,angle=-90]{conf_fit_060729.eps}
}
\caption{Examples fits. The dot-dashed lines show
the individual components. The top panel shows the fitted data, the bottom
panel the hardness ratio. The latter is not fitted,
but is calculated from the top panel, and is shown to illustrate how the
model reproduces the dip feature. Left panel: GRB
061110A. Right panel: GRB 060729}
\end{figure}

\section{Results}

We identified GRBs with known redshift, a clear dip feature in the
hardness ratio, no flares close to or overlapping the dip, and where
Swift observed throughout the dip feature. This gave an list of 23 GRBs.
Histograms of the initial Lorentz factors determined from these fits are
given in Fig.~3. For an ISM-like environment the median value is 225,
with most values being above 100, in agreement with measurements in the
literature (e.g. \cite{Molinari2007}, \cite{Xue2009}, \cite{Kumar2007},
\cite{Melandri2010}). On the other hand, if we assume a wind-like
environment the median value of the initial Lorentz factor is only 93,
and is below 50 for several bursts; this is close to the limit imposed
by the compactness problem  and may suggest that these GRBs cannot have
occurred in a wind-like environment. We note that all of our values are
well below $\Gamma_0=1000$, which has been proposed as a lower limit on
the bulk Lorentz factor given the GeV emission detected by Fermi-LAT
(e.g.\ \cite{Abdo090902B}). However none of the GRBs in our sample are
those which were also observed by the LAT. Also note that \cite{Zhao11}
and Haosco\"et et al.\ (these proceedings) have argued that GeV photons
can escape from a Fireball with bulk Lorentz factors substantially below
1000.

\section{Conclusions}

We have shown that a dip feature seen in the X-ray hardness
ratios of GRBs at the end of the steep decay phase can be interpreted as
the turn-on of the X-ray afterglow, which rises in accordance with the
predictions of \cite{Sari1999}. By modelling this rise we have determined
the initial bulk Lorentz factor for 23 GRBs, finding a median value of
225 assuming an ISM-like circumburst medium, or 93 for a
wind-like medium.


\begin{figure}
  \includegraphics[height=.4\textheight]{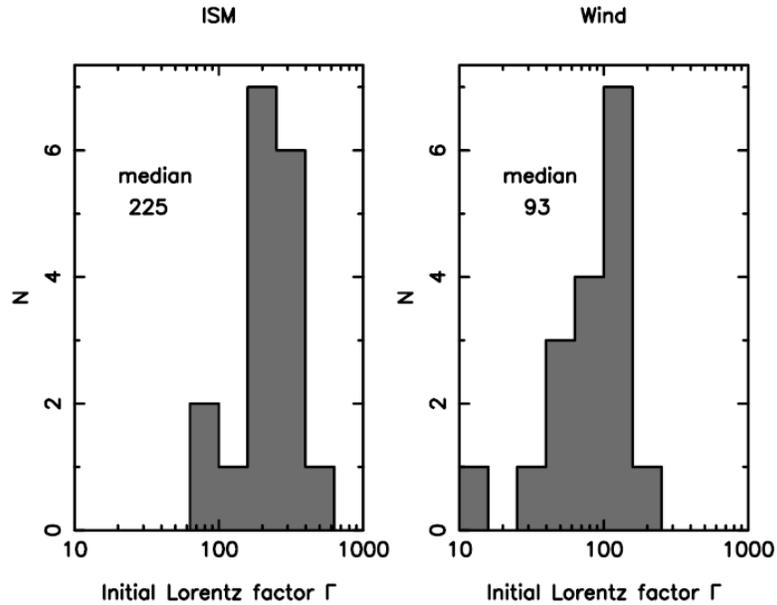}
  \caption{The distribution of fireball initial bulk Lorentz factors
determined from our fitting, for a constant-density (left) or
wind-like (right) circumburst medium.}
\end{figure}




\bibliographystyle{aipproc}   

\bibliography{paper}

\IfFileExists{\jobname.bbl}{}
 {\typeout{}
  \typeout{******************************************}
  \typeout{** Please run "bibtex \jobname" to optain}
  \typeout{** the bibliography and then re-run LaTeX}
  \typeout{** twice to fix the references!}
  \typeout{******************************************}
  \typeout{}
 }

\end{document}